\documentclass[letterpaper,10pt,nofootinbib,aps,tightenlines,twocolumn]{revtex4}
\usepackage{amsmath,amsfonts,amssymb}
\usepackage{mathrsfs}
\usepackage{graphicx}
\usepackage[english]{babel}

\begin{document}

\title{Laboratory search for a quintessence field}
\author{Michael V. Romalis}
\affiliation{Department of Physics, Princeton University, Princeton, New Jersey, 08544,
USA}
\author{ Robert R. Caldwell}
\affiliation{Department of Physics and Astronomy, Dartmouth College Hanover, NH 03755 USA}
\date{\today }

\begin{abstract}
A cosmic scalar field evolving very slowly in time can account for the
observed dark energy of the Universe. Unlike a cosmological constant, an
evolving scalar field also has local spatial gradients due to gravity. If
the scalar field has a minimal derivative coupling to electromagnetism, it
will cause modifications of Maxwell's equations. In particular, in the
presence of a scalar field gradient generated by Earth's gravity, regions
with a magnetic field appear to be electrically charged and regions with a
static electric field appear to contain electric currents. We propose
experiments to detect such effects with sensitivity exceeding current limits
on scalar field interactions from measurements of cosmological
birefringence. The scalar field with derivative couplings to fermions or
photons is also observable in precision spin precession experiments.
\end{abstract}

\maketitle

Many different astrophysical observations point to the existence of dark
energy constituting the majority of the energy density of the Universe \cite%
{Frieman:2008sn}. The origin of this energy is presently unknown, but the
two leading theoretical possibilities are a cosmological constant or a very
slowly evolving dynamical field \cite{Caldwell:2009ix}. These two
possibilities can be distinguished by astrophysical measurements of the
equation of state $w=p/\rho $ -- the ratio of the pressure $p$ to the
density $\rho $ of the dark energy. Present measurements are consistent with
$w=-1$, as expected for a cosmological constant, but also allow $w+1\sim 0.1$
\cite{Hinshaw:2012fq}, as can be obtained in dynamical scalar field models.
Can these two possibilities be distinguished by local measurements, without
relying on astrophysical observations of the dynamical history of the
Universe?

Even though scalar field models are Lorentz covariant, the existence of a
cosmic scalar $\phi$ introduces a special frame in which $\phi$ is
homogeneous. Therefore, any interaction of ordinary matter with the scalar
field can lead to an apparent violation of local Lorentz invariance \cite%
{Pospelov:2004fj,Caldwell:2009zz,Flambaum:2009mz,Kostelecky:2009zp,Arvanitaki:2009fg}%
. If the scalar field is slowly evolving in time, it defines a preferred
cosmic rest frame. Our motion relative to that frame with velocity $v\sim
10^{-3}c $ will lead to small violations of rotational invariance.
However the expected energy shift due to such an effect is on the order $%
v\hbar H_{0}\sim 10^{-45}$~GeV, too small to hope to detect in the
laboratory.

However, a scalar field is also affected by gravity \cite{Caldwell:1997ii}
and will develop a local gradient due to Earth's gravity with a much larger
characteristic energy scale of $\hbar g/c\sim 10^{-31}$~GeV. Such a gradient
in the local value of the scalar field is potentially observable if the
field has any interactions with ordinary matter. In this Letter we explore
possible experimental signatures of such a gravity-induced local scalar
field gradient.

The scalar field equation of motion in the vicinity of a local gravitational
potential, $U$ can be obtained as follows. On length and time scales that
are short compared to the cosmological expansion, we use a weak field
metric:
\begin{equation}
ds^{2}=-(1+2U/c^2)c^2dt^{2}+(1-2U/c^2)d\vec{x}^{2}.
\end{equation}
We consider a static perturbation $\delta \phi$ to the homogeneous solution $%
\phi_0$ in the cosmological background: $\phi =\phi _{0}+\delta \phi (r)$.
Working to linear order, 
and temporarily setting $\hbar=c=1$,
the scalar field equation of motion $\Box \phi =
V^{\prime }$ can be expanded as $\partial_\mu\left( \sqrt{-g}
\,g^{\mu\nu}\partial_\nu\left(\phi_0 + \delta\phi\right) \right) / \sqrt{-g}%
= V^{\prime }(\phi_0+\delta\phi)$ whereupon the background and perturbation
equations are
\begin{eqnarray}
&& \Box_0 \phi_0 = V^{\prime }_0 \cr && \Box_0 \delta\phi -2 U \Box_0\phi_0
= V^{\prime \prime }_0 \delta\phi,
\end{eqnarray}
where the subscript $0$ means that it is evaluated in the unperturbed
spacetime. Using the background solution to simplify the equation for $%
\delta\phi$, for the case of spherical symmetry we obtain
\begin{equation}
\frac{d^{2}}{dr^{2}}\delta \phi +\frac{2}{r}\frac{d}{dr}\delta \phi
=V_{0}^{\prime \prime }\delta \phi +2UV_{0}^{\prime }.
\end{equation}
Using Earth's gravitational potential $U=-GM_{E}/r$, the inhomogeneous
solution is $\delta \phi =-2(V_{0}^{\prime }/V_{0}^{\prime \prime })U/c^2$.
Since the gradient in the gravitational potential gives the local
acceleration, $\mathbf{g}=-\boldsymbol{\nabla }U$, we may write the gradient
in the scalar field as%
\begin{equation}
\boldsymbol{\nabla }\delta \phi =2\frac{V_{0}^{\prime }}{V_{0}^{\prime
\prime }}\mathbf{g}/c^2.  \label{eq:Scalarfieldgrad}
\end{equation}

One of the theoretical challenges to the idea that dark energy is due to a
cosmic scalar field is to explain why the field remains so \textit{dark}.
That is, why doesn't the cosmic scalar mediate a long-range force between
Standard Model particles? One possibility is a global shift symmetry $\phi
\rightarrow \phi ~+~$constant, as occurs in pseudo Nambu-Goldstone boson
models of dark energy \cite{Frieman:1995pm}, which keeps the cosmic scalar
dark \cite{Carroll:1998zi}. The global symmetry allows only derivative
couplings of the scalar field with ordinary matter. For example, consider
the interaction with electromagnetism
\begin{equation}
\mathscr{L}=\frac{1}{2M}\nabla _{\mu }\phi A_{\nu }\widetilde{F}^{\mu \nu }=-%
\frac{\phi }{4M}F_{\mu \nu }\widetilde{F}^{\mu \nu }.  \label{eq:FFdual}
\end{equation}%
In the above, the middle and right terms differ by a total derivative. We
define $\widetilde{F}_{\mu \nu }\equiv \tfrac{1}{2}\epsilon_{\mu \nu \rho
\sigma }F^{\rho \sigma }$, and $M$ is a mass. This coupling introduces a
modification of Maxwell's equations, expressed in SI units where $\phi $ has
units of energy, given by
\begin{eqnarray}
\nabla \cdot \mathbf{E}-\rho /\epsilon_{0} &=&-\frac{1}{Mc}\boldsymbol{%
\nabla }\phi \cdot \mathbf{B},  \label{eqn:ModMaxwell1} \\
\nabla \times \mathbf{B}-\mu _{0}\epsilon_{0}\frac{\partial \mathbf{E}}{%
\partial t}-\mu _{0}\mathbf{J} &=&\frac{1}{Mc^{3}}\left( \dot{\phi}\mathbf{B}%
+\boldsymbol{\nabla }\phi \times \mathbf{E}\right)  \label{eqn:ModMaxwell2}
\end{eqnarray}%
whereas the homogeneous equations are unchanged \cite{Wilczek:1987mv}.

%%%%%%%%%%%%%%%%%%%%%%%%%%%%%%%%%%%%%%%%%%%%%%%%%%%%%%%%%%%

The quintessence field equation is likewise modified,
\begin{equation}  \label{eq:scalarfieldeq}
\square\phi - V^{\prime }(\phi) = \frac{1}{4M}F_{\mu\nu}\widetilde{F}%
^{\mu\nu},
\end{equation}
(again setting $\hbar=c=1$)
where $V(\phi)$ denotes the scalar field potential. For the pseudo
Nambu-Goldstone boson dark energy model, the field self-interacts through a
potential
\begin{equation}
V(\phi) = \mu^4 (1 + \cos(\phi/f)).  \label{eqn:Potential}
\end{equation}

%%%%%%%%%%%%%%%%%%%%%%%%%%%%%%%%%%%%%%%%%%%%%%%%%%%%%%%%%%%

In the vicinity of the Earth we can now rewrite Maxwell's equations as
\begin{eqnarray}
\nabla \cdot \mathbf{E}-\rho /\epsilon_{0} &=&-\frac{\epsilon_{\gamma}}{c}%
\mathbf{g}\cdot \mathbf{B},  \label{eqn:ModMaxwell1b} \\
\nabla \times \mathbf{B}-\mu _{0}\epsilon_{0}\frac{\partial \mathbf{E}}{%
\partial t}-\mu _{0}\mathbf{J} &=&\frac{\epsilon_{\gamma}}{c^{3}}\mathbf{g}%
\times \mathbf{E},  \label{eqn:ModMaxwell2b}
\end{eqnarray}%
where we introduced a dimensionless parameter
\begin{equation}
\epsilon_{\gamma}\equiv \frac{2V_{0}^{\prime }}{V_{0}^{\prime \prime }M c^2}
\end{equation}
and ignored the $\dot{\phi}$-term.

%%%%%%%%%%%%%%%%%%%%%%%%%%%%%%%%%%%%%%%%%%%%%%%%%%%%%%%%%%%

The spatial gradient of the cosmic scalar then introduces novel effects.
Specifically, sources of a magnetic field give rise to an anomalous
electric field, while sources of an electric field will give rise to an anomalous magnetic field.

There is not a unique prediction for $|\epsilon_{\gamma}|$. It could be of
order unity, or tiny with no lower bound. The pseudo-scalar coupling (\ref{eq:FFdual}) causes a
parity-violating rotation of the polarization of light traveling over
cosmological distances in a phenomenon referred to as \textit{cosmic
birefringence} \cite{Carroll:1998zi}. The cosmic microwave background Stokes
vector rotates \cite{Lue:1998mq} by an angle
\begin{equation}
\alpha =\frac{\Delta \phi }{2M c^2},
\end{equation}%
where $\Delta \phi $ is the change in the cosmic scalar between
recombination and the present day. Current bounds, based on the statistical
properties of the polarization pattern detected in the cosmic microwave
background, give $-1.41^{\circ }<\alpha <0.91^{\circ }$ (95\%\thinspace
C.L.), based on a combined analysis \cite{Komatsu:2010fb} of the WMAP \cite%
{Jarosik:2010iu}, BICEP \cite{Chiang:2009xsa,Xia:2009ah}, and QUaD \cite%
{Wu:2008qb,Brown:2009uy} experiments. (Also see Refs.~\cite{Liu:2006uh,Finelli:2008jv} 
for the analysis of particular scalar field models.) The first CMB constraints on a
direction-dependent polarization rotation through cosmological birefringence
has recently been obtained \cite{Gluscevic:2012me}, at a similar amplitude.
For a particular model of quintessence, we can use these observations to
place a lower bound on the mass scale $M$.

\begin{figure}[t]
\includegraphics[width=3.25in]{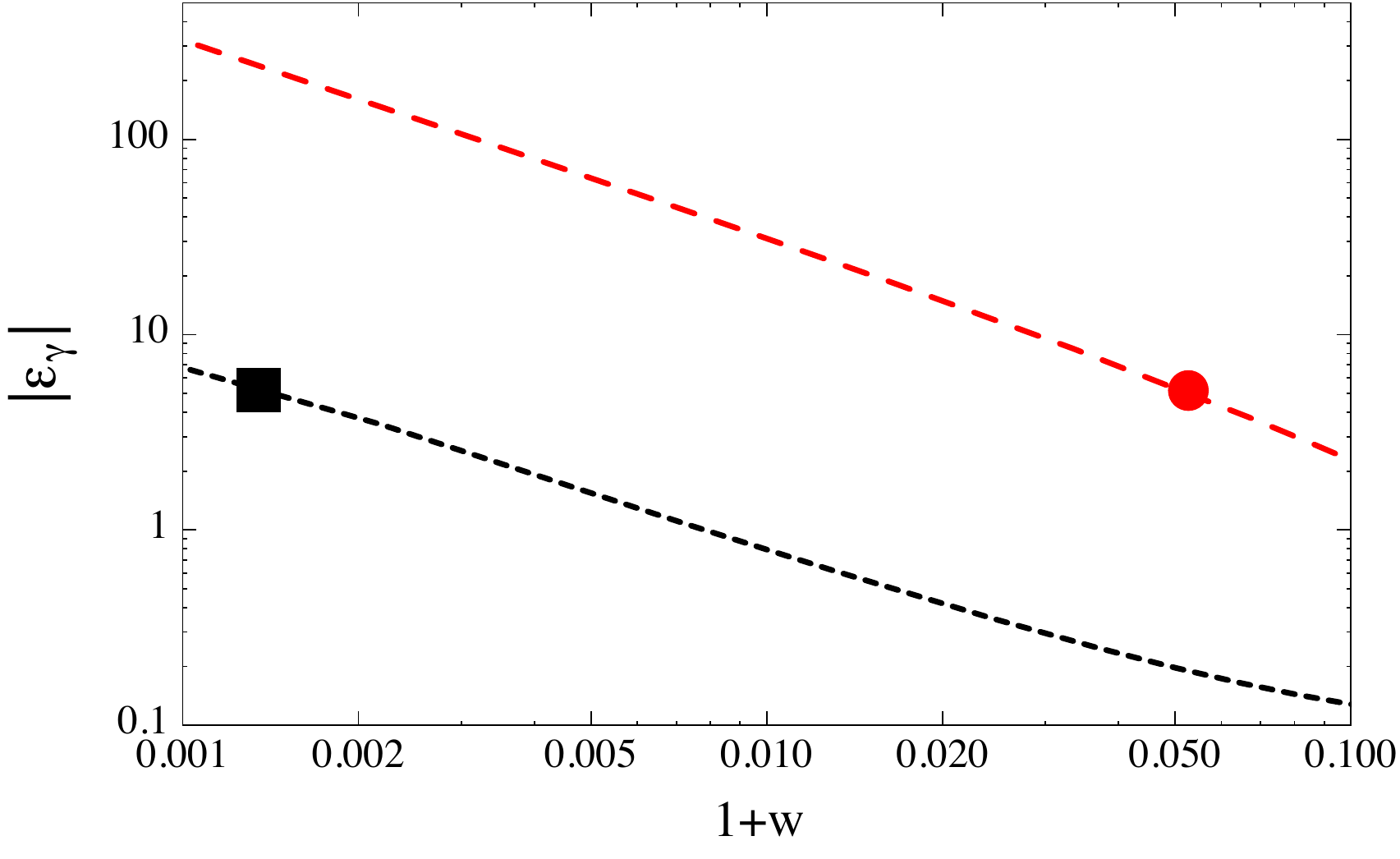}
\caption{The maximum values of $|\protect\epsilon_\protect\gamma|$ versus
the present-day value of the equation of state $w$ for two illustrative
families of pseudo Nambu-Goldstone boson models. For the short (long) dashed
lines, $\protect\mu=0.002 (0.0024)\,{\rm eV/c^2}$. For each point along the curve, $f$
and the initial value of $\protect\phi$ are adjusted so as to yield $%
\Omega_{DE}=0.72$ and $h=0.71$, consistent with WMAP9 \protect\cite%
{Hinshaw:2012fq}. At the square and circle the maximum is $|\protect\epsilon_%
\protect\gamma|=5$. While larger values of $|\protect\epsilon_\protect\gamma|
$ may be achieved with other values of the model parameters, there is no
lower limit on $|\protect\epsilon_\protect\gamma|$.}
\label{fig:PNGB}
\end{figure}

Typical parameters for the pseudo Nambu-Goldstone boson model potential (\ref%
{eqn:Potential}) that satisfy current observational constraints on dark
energy have values $\mu \simeq 0.002~{\rm eV/c^2}$ and $f \sim M_{p}$. The initial
value of the field is not predicted, however, so that in turn the
present-day value of $\phi_0$ is undetermined. Moreover, the field evolution
may have been damped by Hubble friction until quite recently, with a
present-day equation of state close to $-1$. This means that $V^{\prime
\prime }_0$ or $\Delta\phi$ may be small, in which case $|\epsilon_\gamma|$
might conceivably be quite large. Two families of quintessence models for
which the maximum value of $|\epsilon_\gamma|$ ranges over three orders of
magnitude are shown in Fig.~\ref{fig:PNGB}. Models with even larger maxima
can be constructed. For the two models indicated by the square and circle,
the maximum value $|\epsilon_\gamma|$ is $5$. For the model indicated by the
square in the figure, $\mu=0.002\,{\rm eV/c^2}$, $f=0.64\,M_{p}$ with an initial value
$\phi_i=0.38\,M_p$. We assume that the field is frozen by the Hubble
friction, and only begins to slowly roll at late times. From the time of
last scattering to the present, $\Delta\phi = 0.0062\,M_p$, with a current
value $\phi_0 = 0.39\,M_p$ and $w=-0.9985$. For the model indicated by the
circle, $\mu=0.0024\,{\rm eV/c^2}$, $f=0.39\,M_{p}$ and $\phi_i=0.63\,M_p$. From the
time of last scattering to the present, $\Delta\phi = 0.037\,M_p$, with a
current value $\phi_0 = 0.67\,M_p$ and $w=-0.946$. Both models, as well as
the entire families illustrated, are consistent with current cosmological
data. Whereas the model with $w=-0.946$ (circle) may be distinguished from a
cosmological constant by future astrophysical observations, it seems
unlikely that improvements in observations will permit the model with $%
w=-0.9985$ (square) to be distinguished. (See Ref.~\cite{Weinberg:2012es}
for recent forecasts.)

We first consider the experimental consequences of Eq.~(\ref%
{eqn:ModMaxwell1b}). A region of magnetic field will appear electrically
charged, with the sign of the charge depending on whether the magnetic field
is directed up or down relative to Earth's gravity. For a magnetic field $B=1
$~T and $\epsilon_{\gamma}=5$, the effective charge density is $\rho
=1.4\times 10^{-18}$C/m$^{3}=9e$ m$^{-3}$ and can be
detected using fairly conventional techniques. For example, consider an
apparatus shown in Fig.~2. The magnetic field is created by a permanent
Halbach magnet that is rotated around a horizontal axis to modulate the sign
of the effective electric charge inside the cylinder. The charge is detected
by measuring the voltage across a cylindrical capacitor inserted into the
Halbach magnet. The expected unloaded voltage amplitude can be expressed as
\begin{equation}
V=(0.4\;\mathrm{nV})\left( \frac{B}{1T}\right) \left( \frac{D}{\mathrm{20cm}}%
\right) ^{2}\left(\frac{\epsilon_{\gamma}}{5}\right).
\end{equation}%
For optimum detection efficiency the input capacitance of the amplifier
should be equal to the capacitance inside the Halbach magnet, which is about
10 pF for a 30 cm long capacitor. A low-noise JFET transistor with an input
capacitance $C_{in}=2.3$ pF operating at room temperature has a noise level
of $\delta V=4$ nV/Hz$^{1/2}$ at 10 Hz \cite{MOXTEK}. One can use four such
transistors in parallel to match the impedance of the source and reduce
noise. With such an amplifier the signal due to a scalar field with $%
\epsilon_{\gamma}=5$ can be observed at the $4\sigma$ level after 1 hour of
integration.

\begin{figure}[tbp]
\includegraphics[width=2.5in]{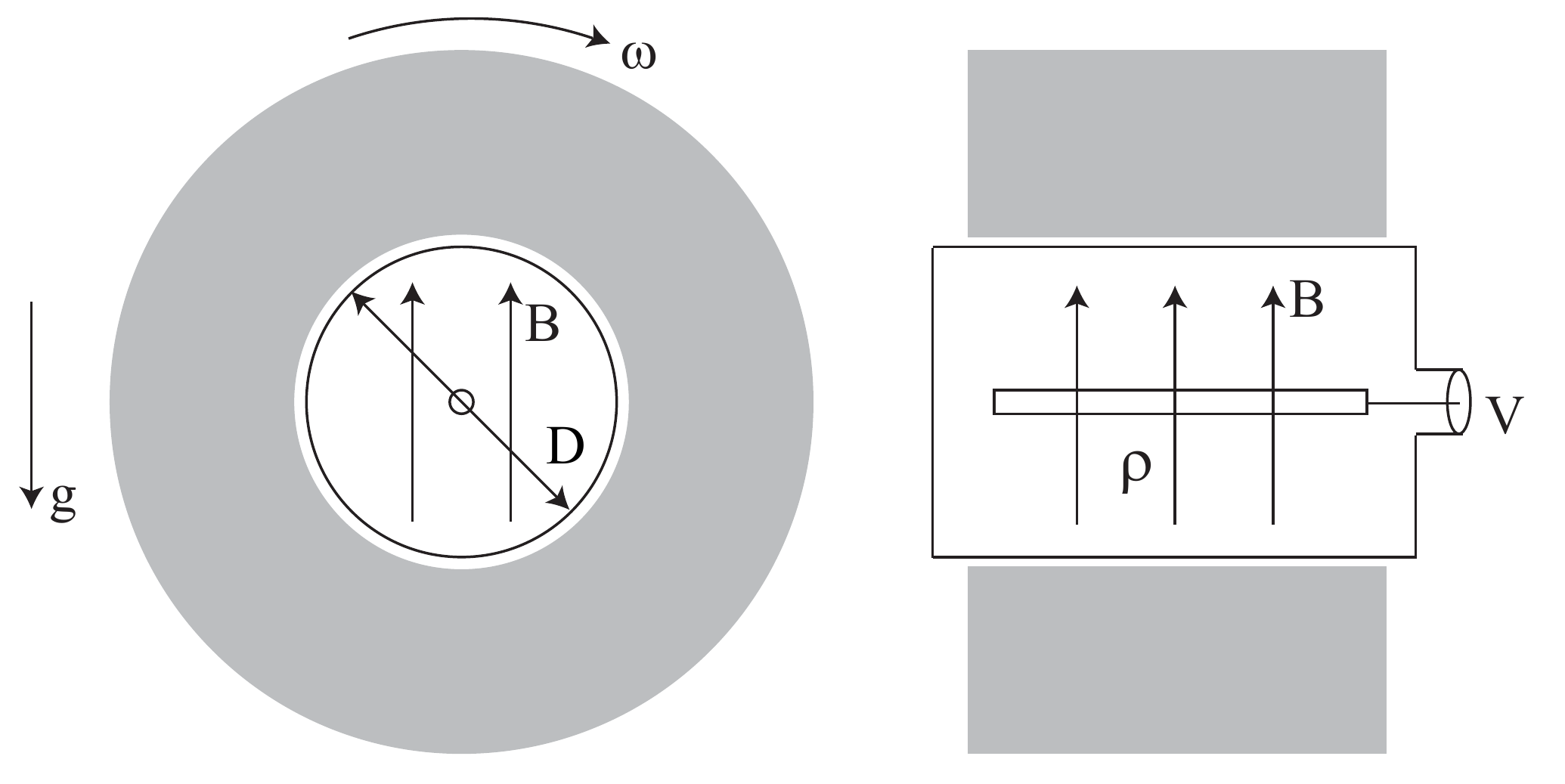}
\caption{Schematic of an experimental setup using a rotating Halbach magnet
to detect modifications of electromagnetism due to a coupling with a
gradient of a scalar field. The voltage $V$ on the cylindrical capacitor
will oscillate with frequency $\protect\omega$.}
\label{fig:Halbach}
\end{figure}

There is considerable room for improvement using more advanced techniques.
Using a superconducting magnet both the magnetic field and the diameter of
the detection region can be increased. The sensitivity could also be
potentially improved using single electron transistors (SET), which can
reach an energy resolution on the order of $\hbar$ \cite{quantumSET}, a
large improvement over the JFET energy resolution of $C_{in}\delta
V^{2}/2=1.8\times 10^{-29}$J~s. However, SETs suffer from $1/f$ noise, so
the best energy resolution demonstrated so far at 10 Hz is $8\times 10^{-30}$%
J~s \cite{lowfreqSET}. In addition, SETs usually have very small input
capacitance, so they would need to be fabricated either with a larger input
gate \cite{largegateSET} or with many of them connected in parallel \cite%
{parallelSET}.

Systematic errors in such an experiment would be primarily due to the
Faraday effect. Even if the capacitor is fabricated to be very axially
symmetric and is rotated together with the magnet, it will inevitably
undergo deformations due to gravity, which will lead to a Faraday induction
signal. However, Faraday signals can be distinguished because they are
proportional to the rotational velocity $\omega$ and reverse sign with the
direction of rotation.

We can also consider an experimental approach to detection of the signal
through Eq.~(\ref{eqn:ModMaxwell2b}) as shown in Fig.~3. The central
grounded cylindrical container is surrounded by electrodes with alternating
high voltage $V$, creating a radial electric field $E$. This field in the
presence of the scalar field gradient generates a circular pattern of
effective current density similar to a solenoid. The effective current
generates an approximately uniform magnetic field inside the central
cylinder, which can be sensed by a magnetometer. Several regions with
alternating field polarity allow one to cancel common magnetic field noise
by using a magnetic gradiometer. The magnetic field in an ideal geometry is
given by
\begin{equation}
B=(1.8\times 10^{-19}\mathrm{T})\left( \frac{V}{\mathrm{100kV}}\right) \left(%
\frac{\epsilon_{\gamma}}{5}\right).
\end{equation}

\begin{figure}[tbp]
\includegraphics[width=1.5in]{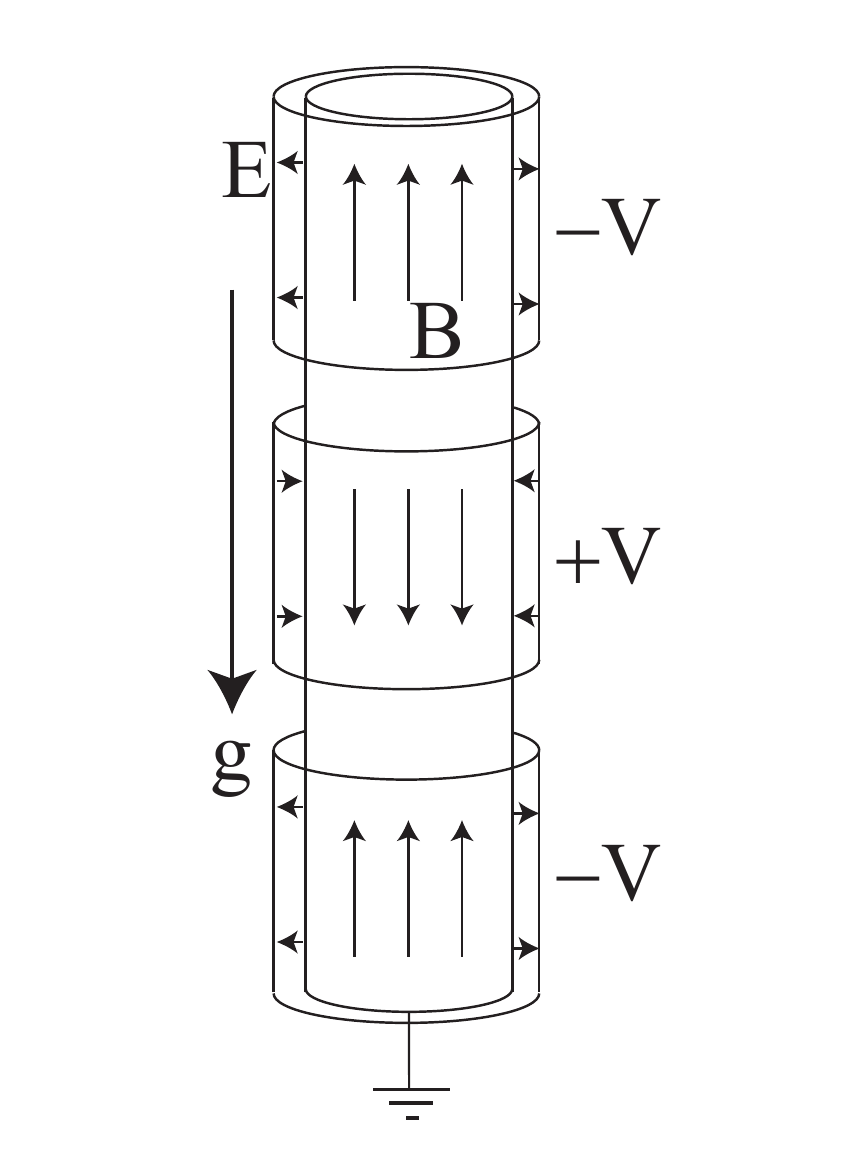}
\caption{Schematic of an experimental setup using an electric field to
generate a magnetic field due to a gradient of the scalar field. A high
voltage $V$ of alternating polarity is applied to electrodes surrounding the
magnetic field sensing region. The sign of the generated magnetic field is
reversed by changing the voltage polarity. }
\label{fig:Emagnet}
\end{figure}

Currently the most sensitive magnetometers using optically-pumped
alkali-metal atoms have a sensitivity of about $\delta B=10^{-16}\mathrm{T/Hz%
}^{1/2}$ for a 1 cm$^{3}$ measurement volume \cite{Dang}. In a long term
measurement such a magnetometer has achieved a sensitivity of $5\times
10^{-19}$ T \cite{Vasilakis}. The sensitivity of atomic magnetometers
improves as $\sqrt{V}$, so with 100 cm$^{3}$ active volume one can achieve a
sensitivity $\delta B=10^{-17}\mathrm{T/Hz}^{1/2}$,  which would allow
detection of the scalar field signal at the $1\sigma$ level after 1 hour of
integration. SQUID magnetometers with a large pick-up coil can also
potentially achieve similar levels of sensitivity \cite{Lamoreaux}.

The experiment can be run similar to an electric dipole moment (EDM)
measurement, in which the electric field polarity is periodically reversed to
modulate the magnetic field. In fact, the recent search for $^{199}$Hg EDM
\cite{EDMsyst} has some sensitivity to $\epsilon _{\gamma }$ coming from $%
^{199}$Hg magnetometer cells maintained at a high potential with no internal
electric fields. While the experiment was not optimized to look for this
effect, analysis of existing data could reach a sensitivity on the order of $%
\epsilon _{\gamma }\sim 3\times 10^{4}$. A dedicated search can reach a much
higher sensitivity since the electric field does not need to be applied
inside the magnetometer cells. Systematic effects in such a measurement
would be similar to an EDM search, primarily due to magnetic fields
generated by charging and leakage currents.

One can consider other ways of detecting such pseudoscalar interactions. The
interaction (\ref{eq:FFdual}) leads to an apparent violation of the equivalence principle \cite{Ni},
but only for spin-polarized bodies since it is proportional to a
pseudoscalar $\mathbf{E\cdot B}$. For example, an electric field around a
nucleus with charge $Ze$ would generate a magnetic field at the origin given by
\begin{equation}
\mathbf{B}=\frac{Ze\epsilon _{\gamma }}{6\pi c^{3}\varepsilon _{0}}\frac{%
\mathbf{g}}{R},
\end{equation}
where $R$ is an integration cut-off which we take to be roughly equal to the
nuclear charge radius \cite{Flambaum:2009mz}. This magnetic field interacts
with the nuclear magnetic moment, causing an effective spin-gravity $\mathbf{S\cdot g}$
frequency shift.  Two experiments \cite{Venema,Peck} have
constrained such  interaction for $^{199}$Hg atoms at a level of
$\Delta \nu <1$ $\mu $Hz, which can be used to place a limit $\epsilon
_{\gamma }\lesssim 3\times 10^{4}$. One can also obtain interesting limits
for electrons from a spin pendulum experiment \cite{Heckel}, which also has $\mu $Hz sensitivity.

Astrophysical sources of gravitational potential can also generate an
observable signal.  The polarization of light escaping from a
gravitational potential which changes by $\Delta U$ will be rotated by an angle
\begin{equation}
\alpha =\frac{\Delta U}{2 c^2}\varepsilon _{\gamma }.
\end{equation}
For example, for Crab nebulae the polarization of gamma rays is measured to
be parallel to the rotation axis of the pulsar within 11$^{\circ }$ \cite{Dean}. 
There is considerable uncertainty in the location where the gamma
rays are  generated \cite{Aharonian}, but assuming a
distance near the light-cylinder radius $R_{L}\sim10^{6}$m
in the potential of a neutron star with $M=1.5M_{\odot }$, one can place a
limit $\varepsilon _{\gamma }\lesssim200$.

In addition to the pseudoscalar derivative coupling to electromagnetism, we
can also consider derivative coupling to fermions, similar to the
interaction discussed, for example, in \cite{Flambaum:2009mz},
\begin{equation}
\mathscr{L}=\frac{1}{2M_{f}}\partial _{\mu }\phi \bar{\psi}\gamma _{\mu
}\gamma _{5}\psi \,.
\end{equation}%
It will lead in the non-relativistic limit to a spin coupling with the
gradient of the scalar field $H=\hbar \boldsymbol{\sigma} \cdot \boldsymbol{\nabla} \phi /(2M_{f}c)$.
Using the gravitationally induced spin gradient from Eq. (\ref{eq:Scalarfieldgrad}), we obtain a frequency shift 
between spin-up and spin-down states equal to
\begin{equation}
\Delta \nu =\frac{g}{2\pi c}\epsilon _{f}=(\mathrm{10nHz})\epsilon _{f},
\end{equation}
where we defined $\epsilon _{f}\equiv 2V_{0}^{\prime }/(V_{0}^{\prime \prime
}M_{f} c^2).$ This frequency shift is similar to the result obtained in \cite{Flambaum:2009mz} 
but without requiring an additional scalar coupling of the
quintessence field to matter, since the scalar gradient is already generated
by ordinary gravity. Atomic physics experiments can easily reach this
frequency resolution \cite{Vasilakis,EDMsyst}, but so far the most sensitive
measurements \cite{Venema,Peck,Heckel} have $\mu $Hz sensitivity,
constraining $\epsilon _{f}<100$. New experiments are being developed to
realize higher sensitivity \cite{Derek}.

In conclusion, we have considered local experimental signatures of a cosmic scalar
field. We point out that Earth's gravity generates a gradient of the
scalar field. Even a minimal derivative coupling of the scalar field to
photons or fermions results in experimentally observable effects. We propose
experiments searching for electromagnetic and spectroscopic signatures of
scalar field interactions. Unlike astrophysical observations, such
experiments could detect the presence of quintessence even if its equation
of state is virtually indistinguishable from a cosmological constant.

\acknowledgments
This work is supported in part by NSF awards PHY-1068027 at Dartmouth
College (RRC) and PHY-0969862 at Princeton (MVR).

\end{document}